\documentclass[superscriptaddress,pra,amsmath,amssymb,twocolumn]{revtex4-2}
\usepackage{graphicx}
\usepackage{float}
\usepackage{dcolumn}
\usepackage{bm}
\usepackage{amsmath,mathrsfs}
\usepackage{graphicx,epsfig}
\usepackage{latexsym,amssymb}
\usepackage{epsf}
\usepackage{color}
\usepackage{hyperref}
\usepackage{setspace}
\usepackage{xcolor}
\usepackage{subcaption} 
\begin{document}
\title{{\bf Curvature-Induced Nonclassicality in a Generalized Jaynes-Cummings Model }}

\author{S. Kourkinejat}\email{s.koorkinejat@gmail.com}
\affiliation{ Physics Department, University of Isfahan, Hezar Jerib St. Isfahan, 81764-73441, Iran.}

\author{ A. Mahdifar }\email{a.mahdifar@sci.ui.as.ir}
\affiliation{ Physics Department, University of Isfahan, Hezar Jerib St. Isfahan, 81764-73441, Iran.}
\affiliation{Quantum Optics Group, Department of Physics, University of Isfahan, Hezar Jerib St. Isfahan, 81746-73441, Isfahan, Iran.}

\author{ E. Amooghorban }\email{ehsan.amooghorban@sku.ac.ir}
\affiliation{Department of Physics,  Faculty of Science,  Shahrekord University P. O. Box 115,  Shahrekord ,  Iran}
\affiliation{Nanotechnology Research Center,  Shahrekord University,  8818634141,  Shahrekord,  Iran}

\date{\today}

\begin{abstract}
In this paper, we investigate the influence of spatial curvature on the Jaynes-Cummings model. We employ an analog model of general relativity, representing the field inside a cavity using oscillators arranged in a circle instead of a straight line, where increasing curvature corresponds to a smaller circle radius.
We investigate the nonclassical features of this quantum system arising from the interaction between a two-level atom and a deformed harmonic oscillator on a circle, which serves as curved-space counterpart to the flat oscillator.
We analyze the time evolution of atom-field states and, based on this dynamic, examine how spatial curvature influences the Mandel parameter, entropy, and the behavior of the Wigner distribution function.
Our results demonstrate that the spatial curvature plays critical roles in controlling these nonclassical properties.
\end{abstract}

\maketitle
\section{Introduction}
For the past decades the Jaynes–Cummings model (JCM) has been playing a very significant role in our understanding of the interaction between radiation and matter in quantum optics.
As one of the most fundamental and analytically solvable models in quantum optics, the JCM describes the interaction between a single-mode quantized electromagnetic field and a two-level atom. The system is modeled by a two-level atom coupled to a harmonic oscillator ~\cite{1443594,PhysRev.140.A1051}. Due to its simplicity and physical relevance, the JCM has become a cornerstone of quantum optics, and its numerous generalizations continue to be the main focus of both theoretical and experimental research. In particular, numerous publications have explored various generalizations of the model in which the interaction between the atom and the radiation field is no longer linear in the field variables  ~\cite{shore01071993,PhysRevA.50.1785,Cordero_2011,PhysRevLett.44.1323,de2011f}.

The representation theory of the quantum algebras with a single deformation parameter $q$ has led to the development of the $q$-deformed oscillator algebra~\cite{n2004dynamical}. The first applications of a $q$-oscillator were to generalize the JCM Hamiltonian with an intensity-dependent coupling by relating it to the quantum suq(1,1) algebra~\cite{PhysRevLett.65.980}. Beyond $q$-deformations, various forms of nonlinearity have been studied in which the oscillator frequency becomes amplitude-dependent, typically characterized by a generic deformation function $f$~\cite{de2011f,BRoy2000,rego2001quantum}. This research path has led to the introduction of $f$-deformed oscillators, which provide a more general framework for modeling nonlinear quantum systems~\cite{man1997f,Mancini:1997em}.

Incorporating physical parameters like the spatial curvature enables us to experimentally probe curvature-induced effects.
The influence of physical space's curvature on certain experiments presents an interesting and significant problem in physics.
In recent years, there have been numerous attempts to employ physical systems, as analogy platforms to simulate and explore some aspects of general relativity theory in laboratory scales. Various experimental efforts also aim to find analogous effects in the lab. These analog models mainly focus on optics~\cite{R16,doi:10.1080/23746149.2020.1759451,PhysRevX.4.011038, doi:10.1080/09500340.2013.769638,Amooghorban2015,Tavakoli2018}. In optics, one can study curved space effects by abandoning one spatial dimension and confining light to a curved surface. Since these models depend on the geometry of physical space, they are regarded as analog models of general relativity~\cite{ R47}

The quantum harmonic oscillator, along with it’s associated coherent states and their generalizations ~\cite{klauder1985coherent}, plays a central role in many theoretical and experimental areas of modern physics, particularly in quantum and atom optics.
Mahdifar and collaborators introduced the coherent states of a harmonic oscillator on the surface of a sphere. These states are explicitly dependent on the curvature of the sphere on which the oscillator oscillates, making them suitable for analyzing curvature-induced effects in physical processes~\cite{mahdifar2006geometric,doi:10.1142/S0219887812500090}.
Recently, Mahdifar and Amooghorban developed coherent states for a harmonic oscillator on a circle, which, like their spherical counterparts, depend on the curvature of the circle on which the oscillator is confined~\cite{doi:10.1142/S0219887822501407}. After  that, Kourkinejat et al. introduced a novel analog model where quantum oscillators on a circle reveal how spatial curvature reshapes blackbody radiation, reducing spectral intensity and width while causing a redshift in the peak radiation frequency~\cite{KOURKINEJAT2025130600}.
Meanwhile, they explores how curvature modulates Planck's distribution, emphasizing geometry's role in thermal light-matter interactions~\cite{kourkinejat2025}.


An essential feature of quantum systems is their nonclassical behavior, which manifests in phenomena such as photon antibunching, sub-Poissonian statistics, entanglement, and negativity in quasiprobability distributions. Mandel was among the first to introduce a quantitative measure for nonclassicality via deviations from the Poissonian photon statistics~\cite{mandel1979sub}. On other hand, entanglement often described as a purely quantum phenomenon, reveals correlations that cannot be explained classically and is deeply tied to measures such as von Neumann entropy.\cite{bianchini2014entanglement, dahl2004interference, ohya2004quantum}. Another powerful tool in assessing quantum nonclassicality is the Wigner function, a quasiprobability distribution in phase space~\cite{carmichael2013statistical}. Although this function originally developed by Wigner, it has attracted renewed attention due to its ability to reveal negative regions, which is a clear sign of non-classical nature~\cite{HUDSON1974249,bell1987speakable}. Later, Kenfack and Zyczkowski introduced a practical criterion of nonclassicality through the volume of the negative part of the Wigner function, which also serves as a useful indicator for quantum entanglement~\cite{kenfack2004negativity,arkhipov2018negativity}.

Inspired by these insights and the emerging signatures of nonclassicality, our work extends this line of inquiry by exploring the impact of spatial curvature on fundamental quantum properties, including photon statistics, entanglement, and phase-space distributions. By incorporating curvature through a deformed algebra applied to a harmonic oscillator on a circle, we generalize the Jaynes-Cummings model to a curved-space framework and investigate how curvature reshapes the system's quantum dynamics and nonclassical features.

The paper is structured as follows:
In Sec.~\ref{JCM}, we briefly review the algebra of a harmonic oscillator on a circle, and focus on the interaction between a quantized field and a two-level atom within the framework of a $\lambda$–dependent Jaynes–Cummings. We study the time evolution of the atomic population inversion and highlight curvature-dependent effects.
Sec.~\ref{DYNAMICAL PROPERTIES}, is devoted to investigating  curvature-induced modifications in the dynamical properties of the atom-field system. We examine the Mandel parameter, the behavior of the field's quantum properties via the generalized Wigner function, and its associated negativity. We also analyze the curvature-dependent entropy as an indicator of entanglement in the system.
Finally, Sec.~\ref{SUMMARY} presents our concluding remarks.
%
\section{ Curvature-Dependent Jaynes-Cummings Model} \label{JCM}
In this section, we adopt an analog model of the general relativity to investigate how the spatial curvature, such as that caused by massive bodies like the Earth, affects properties of quantum optical systems.  For this purpose, we focus on the JCM which describes the interaction between a two-level atom and a quantized field inside a cavity. The quantized electromagnetic field is typically described using creation and annihilation operators, which correspond to oscillators on a straight line. In what follows, we consider a curvature-dependent JCM by modifying its underlying algebra.
\subsection{Curvature-Dependent Oscillator Algebra}
In our analog model, the spatial curvature is incorporated into the atom-field interaction by replacing the flat-space creation and annihilation operators with modified operators corresponding to a harmonic oscillator on a circle of radius $R$, where the spatial curvature is defined as $\lambda={R^{-2}}$. This leads to the formulation of curvature-dependent creation and annihilation operators, as introduced in~\cite{doi:10.1142/S0219887822501407}:
\begin{eqnarray}\label{1}
 \hat{a}_{\lambda} &=& \hat{a} \sqrt{\gamma+ \lambda \frac{\hat{n}-1}{2}},  \\\  \nonumber
 \hat{a}^{\dag}_{\lambda} &=& \sqrt{\gamma+ \lambda \frac{\hat{n}-1}{2}}  \hat{a}^{\dag},
\end{eqnarray}
where  $\hat{n}=\hat{a} ^{\dag}\hat{ a}$  and $ \gamma  = \frac{ (\lambda + \sqrt{{\lambda}^{2} + 4} )}{2}$.
The above $\lambda$-dependent annihilation and creation operators satisfy the following modified commutation relation:
\begin{eqnarray}\label{2}
[\hat{n}, \hat{a}_{\lambda}] &=& - \hat{a}_{\lambda},  \nonumber \\\
[\hat{n}, \hat{a}^{\dag}_{\lambda} ] &=& \hat{a}^{\dag}_{\lambda},   \nonumber \\\
[\hat{a}_{\lambda}, \hat{a}^{\dag}_{\lambda}] &=& \hat{n} \lambda + \gamma.
\end{eqnarray}
In the straight-line limit, $\lambda \rightarrow 0$ (or equivalently $R\rightarrow \infty$), these relations reduce to those of the standard Weyl-Heisenberg algebra.
Furthermore, if we make the identification
\begin{eqnarray}\label{3}
\hat{a}_{\lambda} &\rightarrow & \hat{K}^{-},   \nonumber \\\
\hat{a}^{\dag}_{\lambda} &\rightarrow & \hat{K}^{+},   \nonumber \\\
\frac{\hat{n} \lambda + \gamma}{2} &\rightarrow & \hat{K}^{0},
\end{eqnarray}
we arrive at the $\lambda$-deformed $su(1,1)$ algebra:
\begin{eqnarray}\label{4}
[\hat{K}^{0},\hat{K}^{-}] &=&-\frac{\lambda}{2} \hat{K}^{-}, \nonumber \\\
[\hat{K}^{0},\hat{K}^{+}] &= &\frac{\lambda}{2} \hat{K}^{+},  \nonumber \\\
[\hat{K}^{+},\hat{K}^{-}] &=& -2\hat{K}^{0}.
\end{eqnarray}
With the electromagnetic field described by this curvature-dependent oscillator algebra~\eqref{2}, we are now in a position to formulate the curvature-dependent JCM.
%
\subsection{ Curvature-Induced $\lambda$-Dependent Hamiltonian}
The interaction of an atom (or an ensemble of atoms) with a quantized field and it’s physical realization as cavity quantum electrodynamics provides a suitable system and platform to examine various quantum principles ~\cite{haroche2006exploring}.
Using the framework of quantum harmonic oscillators on a circle, we extend the conventional Jaynes–Cummings model to incorporate the effects of spatial curvature. Within this generalized analog model, referred to the $\lambda$-dependent Jaynes-Cummings model, the quantized field is described by oscillators on a circle. Under the rotating wave approximation, the corresponding Hamiltonian takes the form:
\begin{equation}\label{5}
 \hat{H}_{JC}^{\lambda} = \hbar \omega \hat{a}_{\lambda}^{\dag} \hat{ a}_{\lambda} + \frac{\hbar}{2}\omega_{eg} \hat{\sigma}_{z} + \hbar g \Big(\hat{a}_{\lambda}^{\dag} \hat{\sigma }^{-}+\hat{ a}_{\lambda}{\hat{\sigma}}^{\dag} \Big),
\end{equation}
where  the two atomic levels separated by an energy difference $\hbar \omega_{eg}$ are represented by the Pauli matrices $\hat{\sigma}_{z}$ and $\sigma^{\pm}$,  $\omega$ is the single mode field frequency and the atom–field coupling strength is measured by the positive parameter $g$.
In order to study the influence of the spatial curvature on the dynamics of the system, firstly it is convenient to split up the Hamiltonian  \eqref{5}, as:
\begin{equation}\label{6}
 \hat{H}_{JC}^{\lambda}= \hat{H}_{0}^{\lambda}+ \hat{H}_{1}^{\lambda},
\end{equation}
where,
\begin{equation}\label{7}
 \hat{H}_{0}^{\lambda}= \hbar \omega \hat{a}_{\lambda}^{\dag} \hat{ a}_{\lambda} + \frac{1}{2}\omega_{eg} \hat{\sigma}_{z},
\end{equation}
and
\begin{equation}\label{8}
\hat{H}_{1}^{\lambda} = \hbar g \Big(\hat{a_{\lambda}}^{\dag} \hat{\sigma }^{-}+\hat{ a}_{\lambda}{\hat{\sigma}}^{\dag} \Big).
\end{equation}
In the interaction picture with respect to $ \hat{H}_{0}^{\lambda}$, the interaction Hamiltonian of the system becomes:
\begin{eqnarray}\label{9}
\hat{H}_{I}^{\lambda}&=& e^{\frac{i \hat{H}_{0}^{\lambda} t}{\hbar}} \hat{H}_{1}^{\lambda}  e^{-\frac{i \hat{H}_{0}^{\lambda} t}{\hbar}},\\
&=&\hbar g \Big[\hat{\sigma}_{+} e^{i\Omega^{\lambda}_{\hat{n}} t} \hat{ a}_{\lambda} - \hat{a}_{\lambda}^{\dag} e^{- i\Omega^{\lambda}_{\hat{n}} t} \hat{\sigma }_{-} \Big],\nonumber
\end{eqnarray}
where  $\Omega^{\lambda}_{\hat{n}} = \omega_{eg} - \omega \big(\hat{n} \lambda + \gamma \big) $ can be interpreted as a $\lambda$-dependent detuning.
\subsection{ Curvature-Influenced Atom-Field Dynamics}
In order to study the dynamics of the system governed by the $\lambda$-dependent Hamiltonian~\eqref{9}, we should solve the time-dependent schrodinger equation:
\begin{equation}\label{11}
i\hbar \frac{\partial}{\partial t} |\Psi^{\lambda}(t) \rangle = \hat{H}^{\lambda}_{I} |\Psi^{\lambda}(t) \rangle.
\end{equation}
The interaction Hamiltonian \eqref{8} dictates that the state $|e,n \rangle$ evolves into $|g,n+1 \rangle$ and vice versa. This implies that the entire Hilbert space decomposes into subspaces spanned by  $|e,n \rangle$ and $|g,n+1 \rangle$, ensuring that the dynamics are confined to these individual subspaces. Therefore, the atom-field state at any time $t>0$ can be written as:
\begin{equation}\label{12}
|\Psi^{\lambda}(t) \rangle = \sum^{\infty}_{n=0} \Big[c^{\lambda}_{e.n}(t) |e,n \rangle + c^{\lambda}_{g.n+1}(t) |g,n +1 \rangle\Big].
\end{equation}

By inserting the atom–field state \eqref{12} into the schrodinger equation \eqref{11} and using the Hamiltonian \eqref{9}, the probability amplitudes $c^{\lambda}_{e,n}(t)$ and $c^{\lambda}_{g,n+1}(t)$ are obtained from the following coupled equations:
\begin{equation}\label{13}
\dot{c}^{\lambda}_{e,n}(t) = -ig \sqrt{(\gamma+ \lambda \frac{n}{2})(n+1)}~ e^{i\Omega^{\lambda}_{n} t}~c^{\lambda}_{g,n+1}(t),
\end{equation}
and
\begin{equation}\label{14}
\dot{c}^{\lambda}_{g,n+1} (t)= -ig \sqrt{(\gamma+ \lambda \frac{n}{2})(n+1)}~ e^{-i\Omega^{\lambda}_{n} t}~c^{\lambda}_{e,n}(t).
\end{equation}
Solving Eqs.~\eqref{13} and \eqref{14}, the general expressions for the probability amplitudes are obtained as:
\begin{eqnarray}\label{15}
\hspace{-0.5cm} c^{\lambda}_{e,n}(t) &=& c^{\lambda}_{e,n}(0) \Big[\cos(\frac{\Phi^{\lambda}_{n}t}{2}) - \frac{i\Omega_{n}^{\lambda}}{\Phi_{n}^{\lambda}}\sin(\frac{\Phi^{\lambda}_{n}t}{2}) \Big] e^{\frac{i\Omega^{\lambda}_{n} t}{2}}  \nonumber\\
  &-&  c^{\lambda}_{g,n+1}(0)~\frac{2 i g \sqrt{(\gamma+ \lambda \frac{n}{2})(n+1)}}{\Phi^{\lambda}_{n}} \sin (\frac{\Phi^{\lambda}_{n}t}{2})~ e^{\frac{i\Omega^{\lambda}_{n} t}{2}}, \nonumber \\
\hspace{-0.5cm} c^{\lambda}_{g,n+1}(t) &=& c^{\lambda}_{g,n+1}(0) \Big[\cos(\frac{\Phi^{\lambda}_{n}t}{2})+ \frac{i\Omega^{\lambda}_{n}}{\Phi^{\lambda}_{n}}\sin(\frac{\Phi^{\lambda}_{n}t}{2}) \Big] e^{-\frac{i\Omega^{\lambda}_{n} t}{2}}  \nonumber\\
  &-& c^{\lambda}_{e,n}(0)~\frac{2 i g \sqrt{(\gamma+ \lambda \frac{n}{2})(n+1)}}{\Phi^{\lambda}_{n}} \sin (\frac{\Phi^{\lambda}_{n}t}{2})~ e^{-\frac{i\Omega^{\lambda}_{n} t}{2}},  \nonumber \\
\nonumber \\
\end{eqnarray}
where $\Phi^{\lambda}_{n}$ is the curvature-dependent Rabi frequency defined by:

\begin{equation}\label{16}
\Phi^{\lambda}_{n}= \sqrt{(\Omega^{\lambda}_{n})^{2}+ 4 g^{2} (n+1)(\gamma+ \lambda \frac{n}{2})}.
\end{equation}
It is obvious that in the flat limit ($\lambda\rightarrow 0$), Eq.~\eqref{16} reduces to the standard Jaynes–Cummings Rabi frequency $\Phi^{\lambda=0}_{n}= \sqrt{\Delta ^{2}+ 4 g^{2}(n+1)}$ with detuning $ \Delta = \omega-\omega_{eg} $.

In the following we assume that the atom is initially prepared in the excited state $| e \rangle$, so that we have: $c^{\lambda}_{e,n}(0)= c^{\lambda}_{n}(0)$ and $c^{\lambda}_{g,n+1}(0)=0$, where the initial state of the field is described by $c^{\lambda}_{n}(0)$. Therefore, Eq.\eqref{15} can be simplified as follows:
\begin{eqnarray}\label{17}
c^{\lambda}_{g,n+1}(t) &=& - c^{\lambda}_{n}(0) \frac{2 i g \sqrt{(\gamma+ \lambda \frac{n}{2})(n+1)}}{\Phi^{\lambda}_{n}}  \sin (\frac{\Phi^{\lambda}_{n}t}{2}) e^{-\frac{i\Omega^{\lambda}_{n} t}{2}}, \nonumber \\\
c^{\lambda}_{e,n}(t) &=& c^{\lambda}_{n}(0)\Big[\cos(\frac{\Phi^{\lambda}_{n}t}{2}) - \frac{i\Omega^{\lambda}_{n}}{\Phi^{\lambda}_{n}}\sin(\frac{\Phi^{\lambda}_{n}t}{2})\Big] e^{\frac{i\Omega^{\lambda}_{n} t}{2}}.
\end{eqnarray}
Furthermore, if we assume that the initial field  is prepared in a coherent state:
\begin{equation}\label{18}
|\alpha\rangle = e^{-\frac{|\alpha|^{2}}{2}}\sum^{\infty}_{n=0} \frac{\alpha^{n}}{\sqrt{n!}} |n\rangle,
\end{equation}
 we have:
 \begin{equation}\label{19}
c^{\lambda}_{n}(0) = e^{-\frac{|\alpha|^{2}}{2}} \frac{\alpha^{n}}{\sqrt{n!}}.
\end{equation}
In the next section, we analyze the influence of curvature on the dynamical properties of this analog JCM.
%
\section{CURVATURE EFFECTS ON THE DYNAMICAL PROPERTIES OF THE ATOM}\label{ATOM}
 The revival of atomic inversion is widely recognized as a result of quantum interference. These nonclassical effects arise from quantum coherences that develop during the interaction between the atom and the field~\cite{yoo1981non,Naderi2011}.
At any time $t\neq0$, the probabilities of finding  the atom in the states $|e\rangle$ and $| g \rangle$ are given by  $|c^{\lambda}_{e,n}(t)|^{2}$ and  $|c^{\lambda}_{g,n+1}(t)|^{2}$, respectively. Therefore, the atomic population inversion can be written as:
 \begin{equation}\label{20}
\langle \Psi^{\lambda}(t) | \hat{\sigma}_{z} | \Psi^{\lambda}(t)\rangle= |c^{\lambda}_{e,n}(t)|^{2} - |c^{\lambda}_{g,n+1}(t)|^{2}.
\end{equation}
Substituting $c^{\lambda}_{e,n}(t)$ and $c^{\lambda}_{g,n+1}(t)$ from Eq.~\eqref{17} into Eq.~\eqref{20}, we arrive at the following expression for the atomic inversion:
 \begin{eqnarray}\label{21}
\hspace{-0.3cm} \langle \Psi^{\lambda}(t) | \hat{\sigma}_{z} | \Psi^{\lambda}(t)\rangle &= &\sum^{\infty}_{n=0} |c^{\lambda}_{n}(0)|^{2} \\
&\times&\Big[(\frac{\Omega^{\lambda}_{n}}{\Phi^{\lambda}_{n}})^{2}-\frac{4 g^{2}(n+1) (\gamma+ \lambda \frac{n}{2})} {({\Phi}_{n}^{\lambda})^{2}}\cos(\Phi^{\lambda}_{n}t) \Big].\nonumber
\end{eqnarray}
\begin{figure}
\centering
\includegraphics[width=3 in]{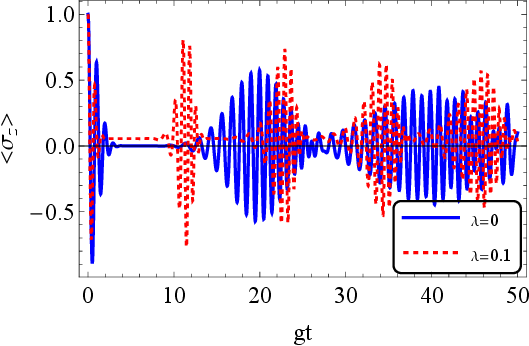}
\caption{  Time evolution of the population inversion when the
field initially prepared in a coherent state with $\alpha=3$. The solid
blue corresponds to $\lambda=0$, and the dotted red to $\lambda=0.1$.}
 \label{fig:1}
\end{figure}
In Fig. \ref{fig:1},  we have plotted the atomic inversion as a function of the scaled time $gt$, at the exact resonance $(\Delta = 0)$,  for different values of spatial curvature.  The comparison of the atomic inversion in flat space $(\lambda=0)$ and that in curved space $(\lambda=0.1)$, highlights the effect of the spatial curvature on the population inversion.
It is observed that with increased spatial curvature, the revivals appear at shorter intervals.
One can estimate the time intervals between successive revivals of the curvature-dependent Rabi oscillations. The occupation probabilities consist of a sum of oscillating terms, each term oscillating at a particular Rabi frequency $\Phi^{\lambda}_{n}$. When two neighboring terms oscillate $\pi$ out of phase, they approximately cancel each other. Conversely, constructive interference occurs when neighboring terms are in phase, i.e., when their phase difference is a multiple of $2\pi$. Since only terms near the mean photon number $\langle\hat{ n} \rangle$ contribute significantly, the time interval between revivals is obtain by:
\begin{equation}\label{22}
(\Phi^{\lambda}_{\langle \hat{n} \rangle +1} - \Phi^{\lambda}_{\langle \hat{n} \rangle}) t_{r} =2 \pi m.
\end{equation}
In the limit  $\langle n \rangle \gg 1$, we obtain:
\begin{equation}\label{23}
t_{r} = \frac{2 \pi m \big( \Phi^{\lambda}_{\langle \hat{n} \rangle+1}+\Phi^{\lambda}_{\langle \hat{n} \rangle}\big)}{(\Phi^{\lambda}_{\langle \hat{n} \rangle +1})^{2}-(\Phi^{\lambda}_{\langle \hat{n} \rangle})^{2}}.
\end{equation}
\begin{figure}
\centering
\includegraphics[width=2.8 in]{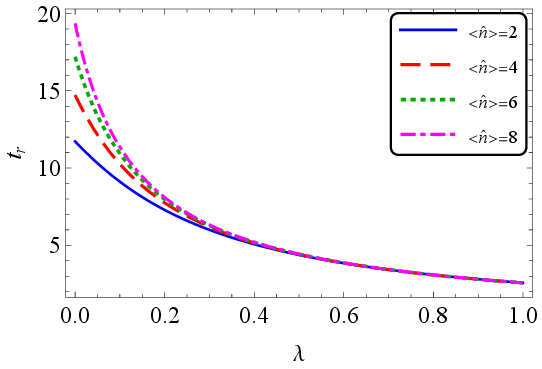}
\caption{Dependence of $t_{r}$ on the curvature $\lambda$,  for different value $\langle \hat{n} \rangle$. The solid blue curve corresponds $\langle \hat{n} \rangle =2$, the dashed red curve to $\langle \hat{n} \rangle = 4$, the dotted green curve to $\langle n \rangle = 6$ and the dotted-dashed magenta curve to $\langle \hat{n} \rangle = 8$.}
 \label{fig:2}
\end{figure}
Fig. \ref{fig:2} illustrates how the revival time $t_{r}$ varies with spatial curvature $\lambda$ for several values of $\langle \hat{n} \rangle$. We see that by increasing the mean number of photons the time interval $t_{r}$ is increased, while increasing the spatial curvature $\lambda$, results in shorter revival intervals. This implies that the revival time of the Rabi oscillations of the atomic occupation probabilities near a massive body would be less than that far away from the massive body ~\cite{mahdifar2013curvature}. This finding is consistent with the time dilation (or gravitational redshift) observed near massive bodies in general relativity~\cite{stephani2004relativity}.
\section{CURVATURE EFFECTS ON THE DYNAMICAL PROPERTIES OF THE FIELDS}\label{DYNAMICAL PROPERTIES}
In this section, we explore how spatial curvature influences the statistical and dynamical behavior of the field. By analyzing photon-counting statistics, the Wigner function, and the atomic entropy, we aim to highlight the nonclassical features that arise in a curved-space framework. These features exhibit notable deviations from flat-space behavior, revealing how curvature modifies the underlying nonclassical properties.
\subsection{Photon counting statistics}
\begin{figure}
\centering
\includegraphics[width=2.8 in]{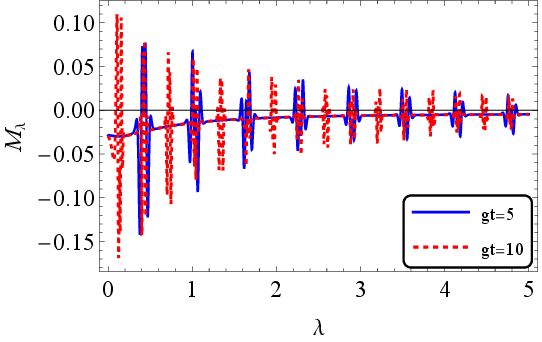}
\caption{ The time evolution of the curvature–dependent Mandel parameter versus the curvature $\lambda$ for  $\alpha=3$ and $\Delta=0$. The solid blue curve corresponds $gt=5$, and the dashed red curve to $gt=10$.}
 \label{fig:3}
\end{figure}

\begin{figure}
\centering
\includegraphics[width=2.8 in]{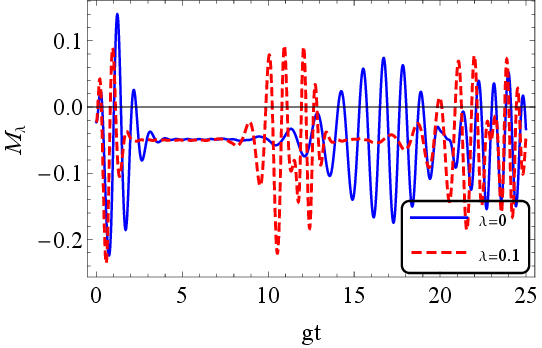}
\caption{ The time evolution of the curvature–dependent Mandel parameter versus the dimensionless, $gt$, with $\alpha=3$. The solid blue curve corresponds $\lambda= 0$, and the dashed red curve to $\lambda=0.1$. }
 \label{fig:4}
\end{figure}

In order to measure the departure of the photon statistics from the Poissonian statistics  of coherent field, we can evaluate the Mandel parameter ~\cite{mandel1979sub,Mandel1986}:
\begin{equation}\label{24}
M= \frac{\langle \hat{n}^{2}  \rangle  - \langle \hat{n}  \rangle ^{2}}{\langle \hat{n}  \rangle} -1,
\end{equation}
where a negative value for $M$ indicates a sub-Poissonian distribution, while a positive value refers to a super-Poissonian distribution. In Fig. \ref{fig:3}, we have plotted  the curvature–dependent Mandel parameter, $M_{\lambda}$, with respect to $\lambda$ for various values of the dimensionless parameter $gt$. Furthermore, Fig. \ref{fig:4} displays  $M_{\lambda}$ versus $gt$ for different curvature values. As observed, the Mandel parameter exhibits a collapse-revival pattern over time, reflecting nonclassical dynamics in the photon statistics. In regions of periodic oscillation, the Mandel parameter $M_{\lambda}$ alternates between positive and negative values, indicating that the field transitions between super-Poissonian and sub-Poissonian statistics at different times.
Furthermore, the most negative value of the Mandel parameter appears at the initial steps of the atom–field interaction.
We also observe that the most negative Mandel parameter occurs at lower curvature values, suggesting that strong curvature may suppress nonclassical statistical features of the field.

\begin{figure*}[t]
\includegraphics[width=0.7\columnwidth]{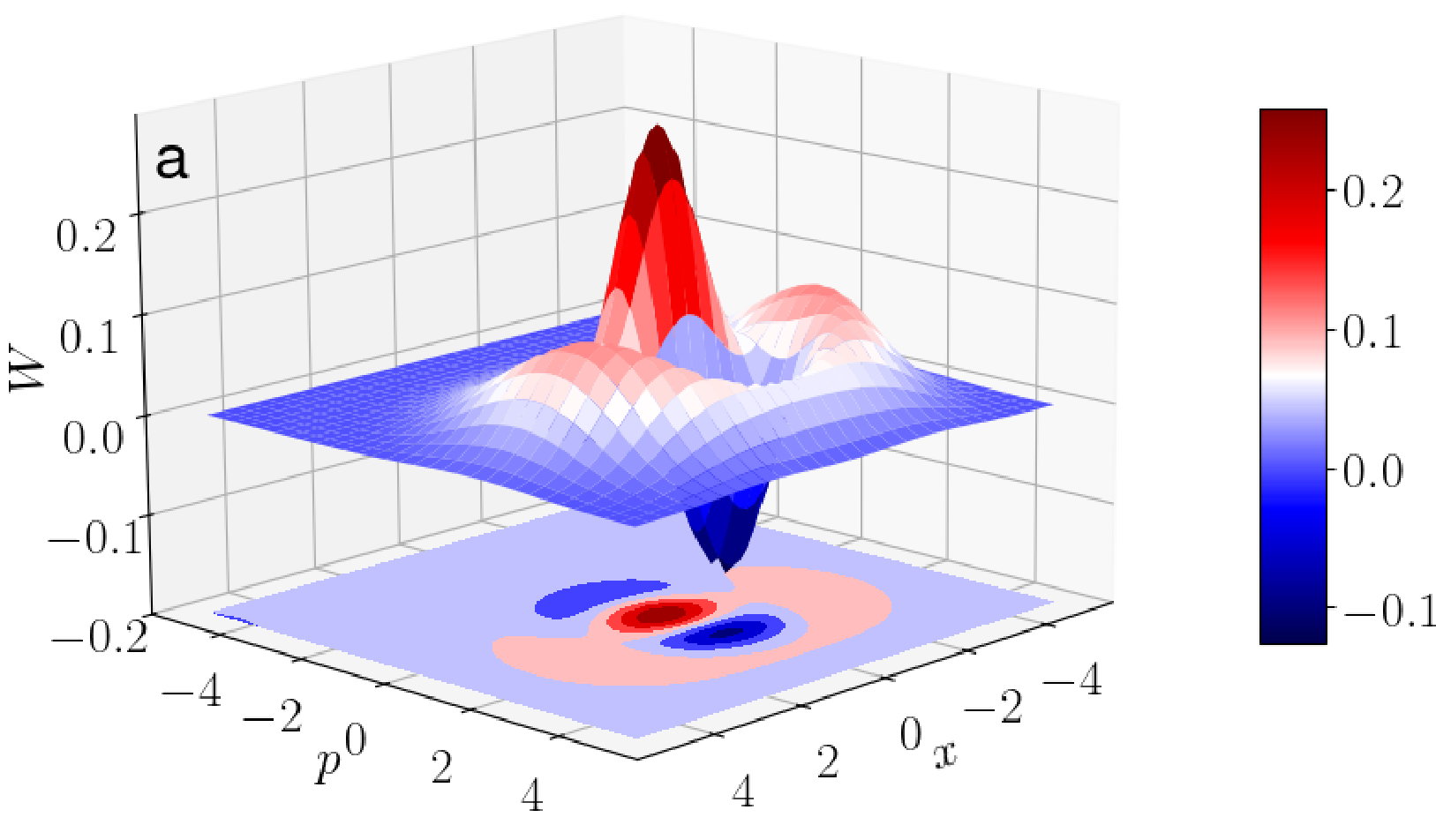}%
\includegraphics[width=0.7\columnwidth]{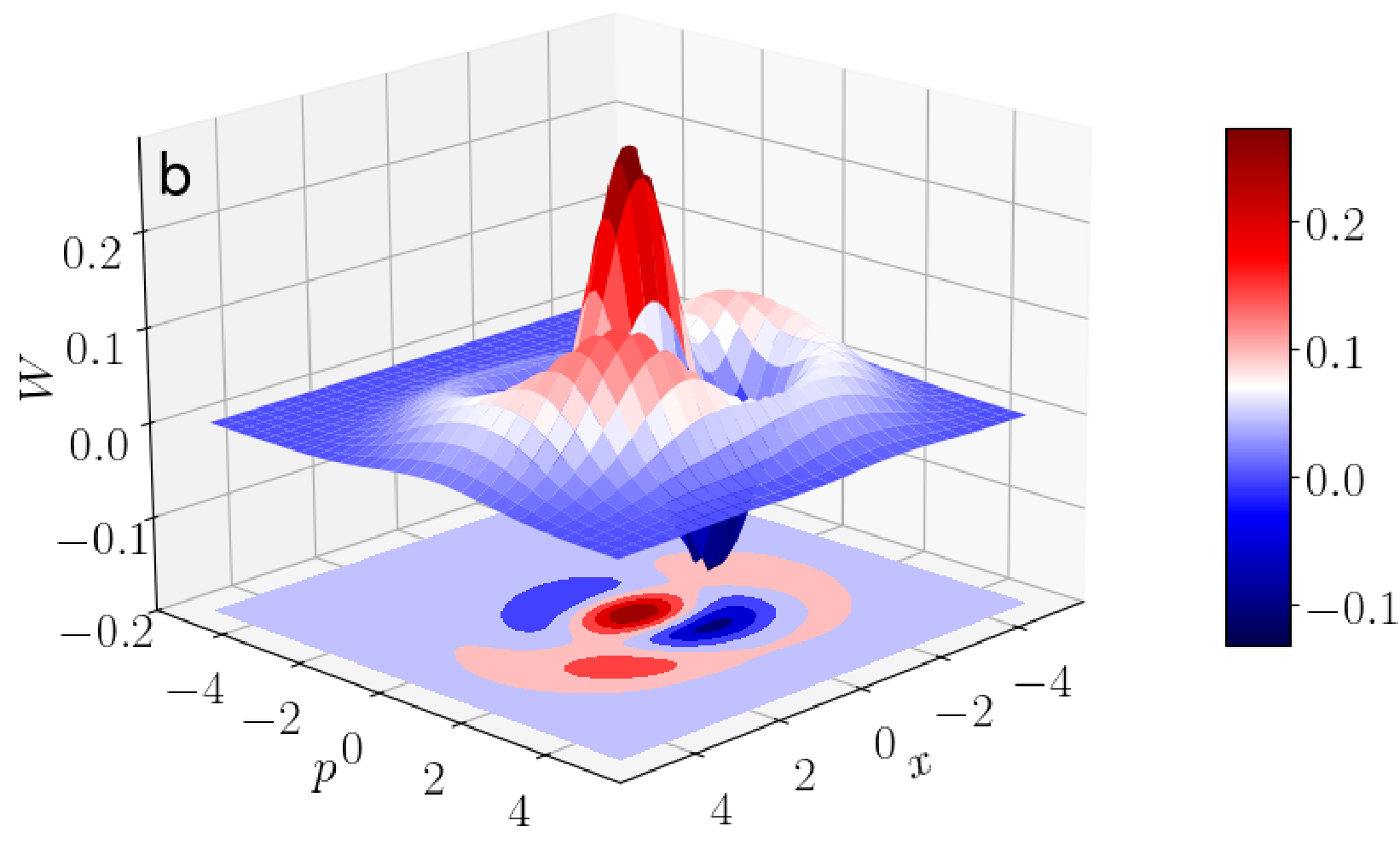}%
\includegraphics[width=0.7\columnwidth]{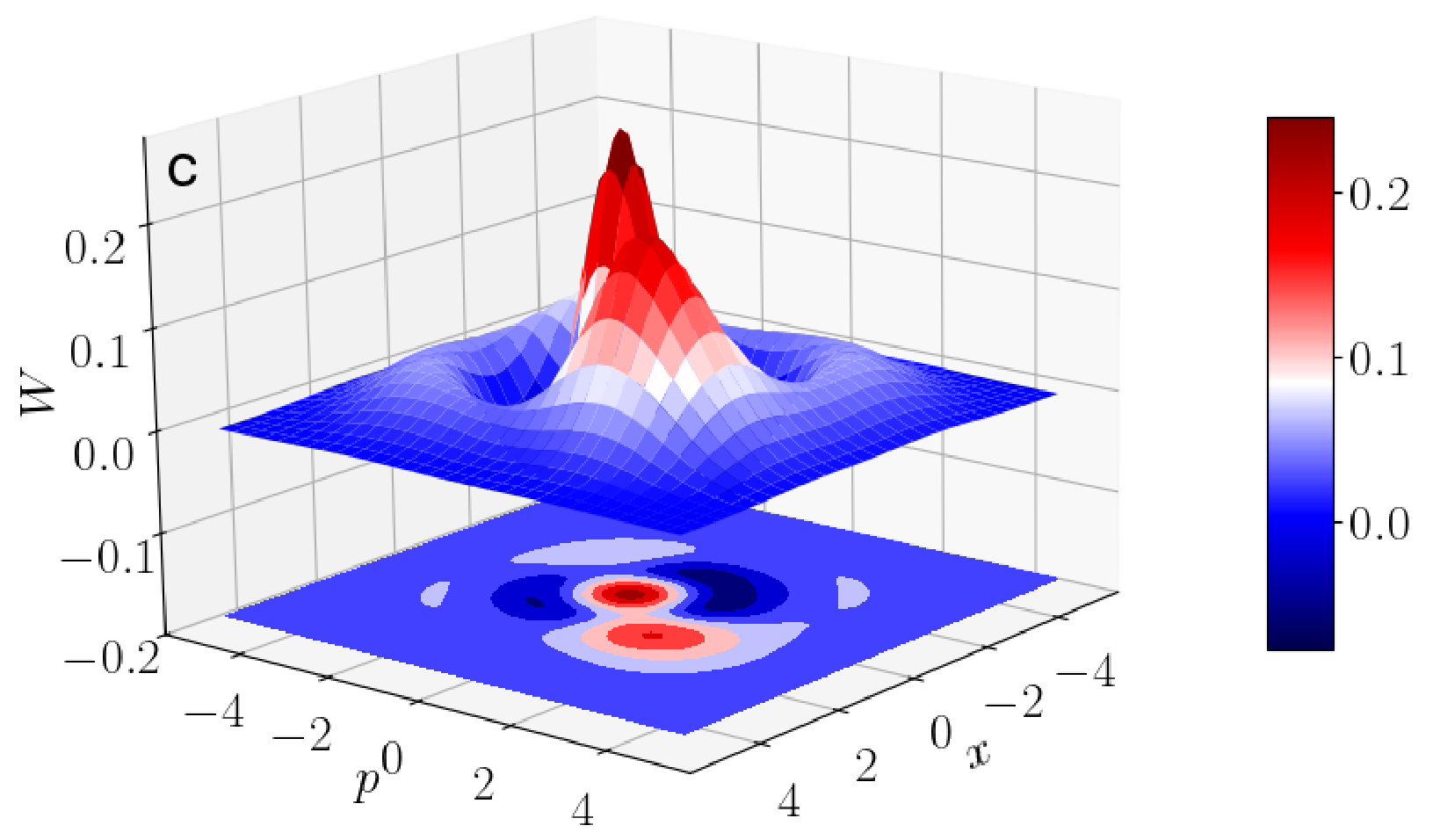}%
\\
\includegraphics[width=0.7\columnwidth]{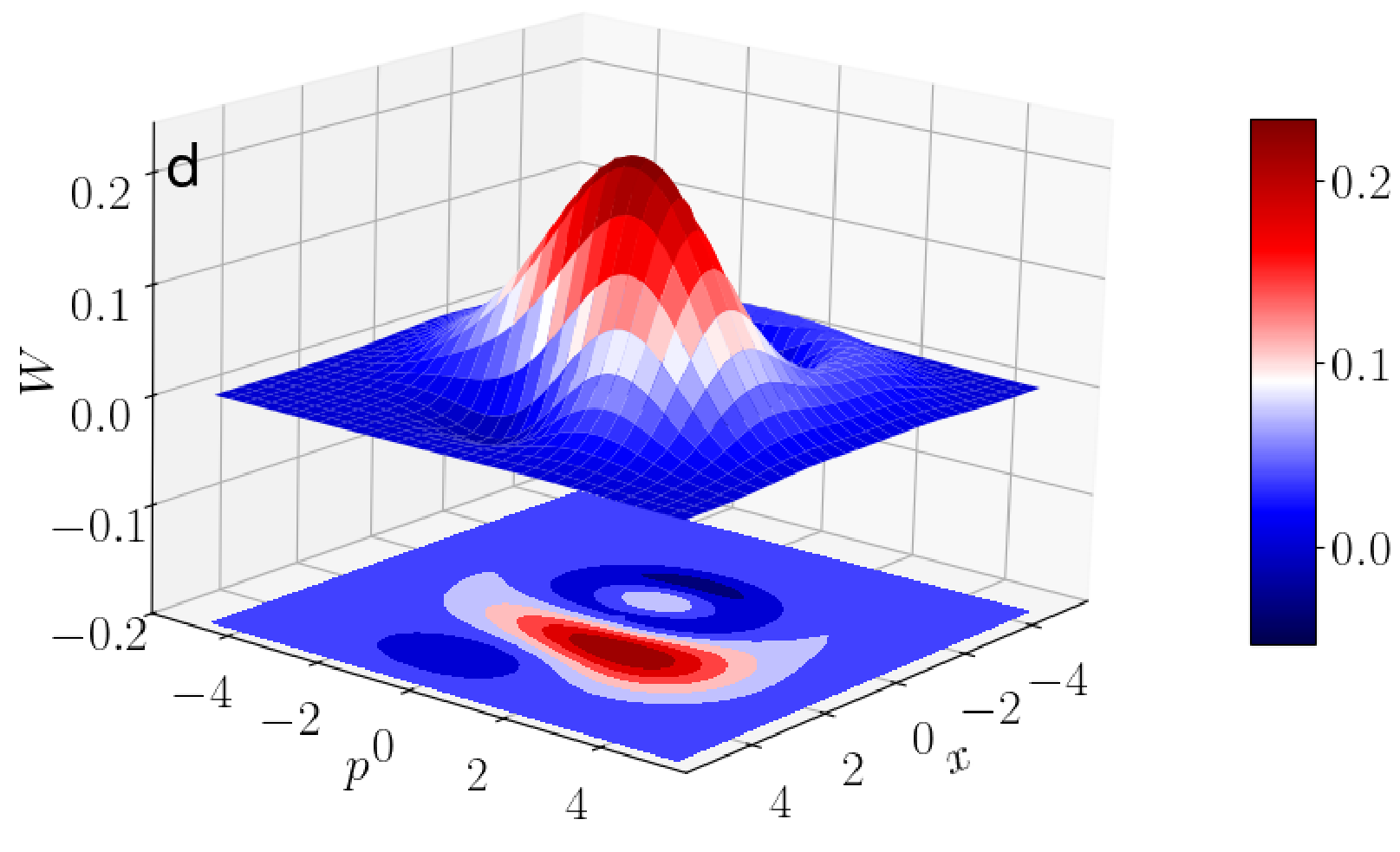}%
\includegraphics[width=0.7\columnwidth]{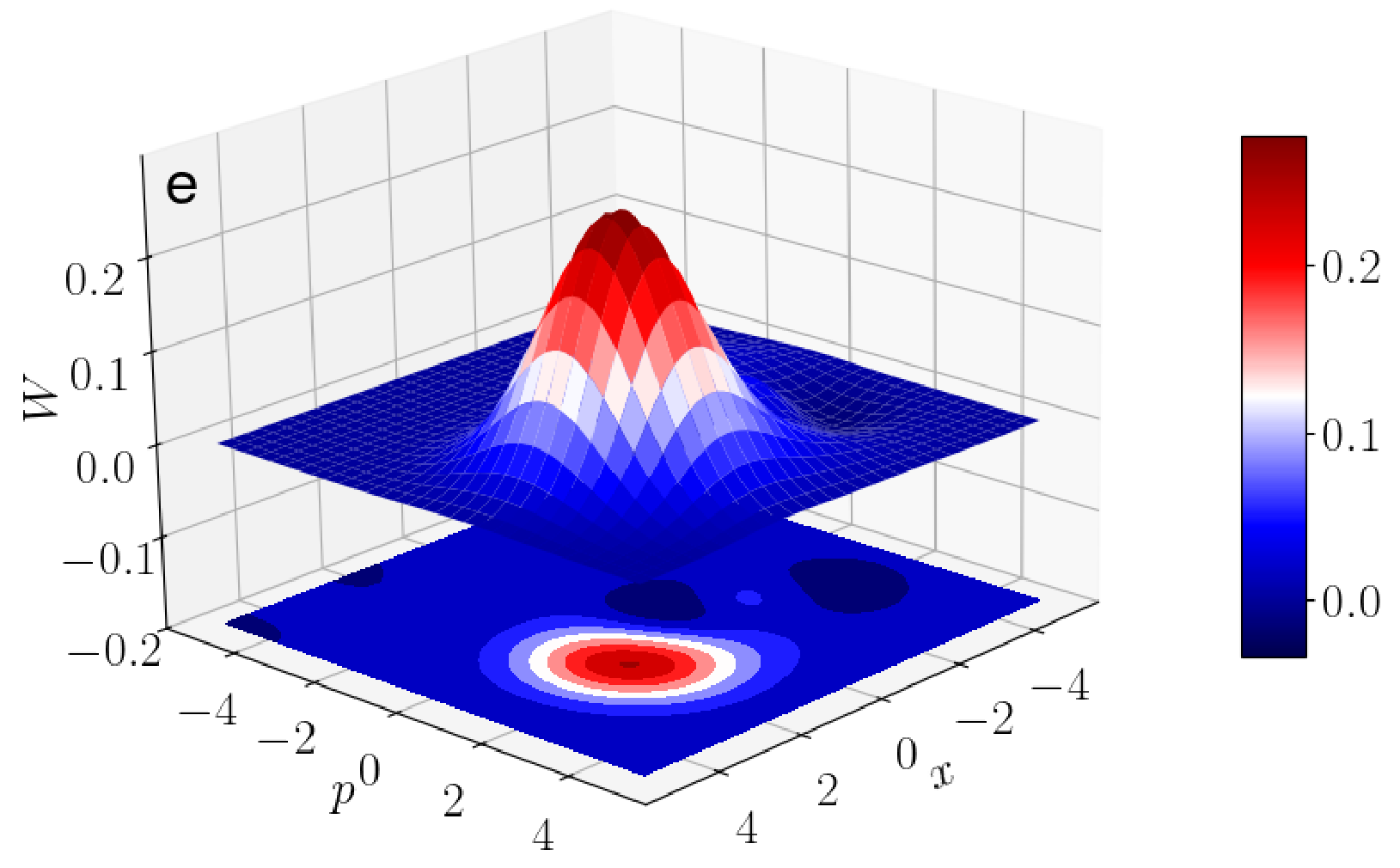}%
\includegraphics[width=0.7\columnwidth]{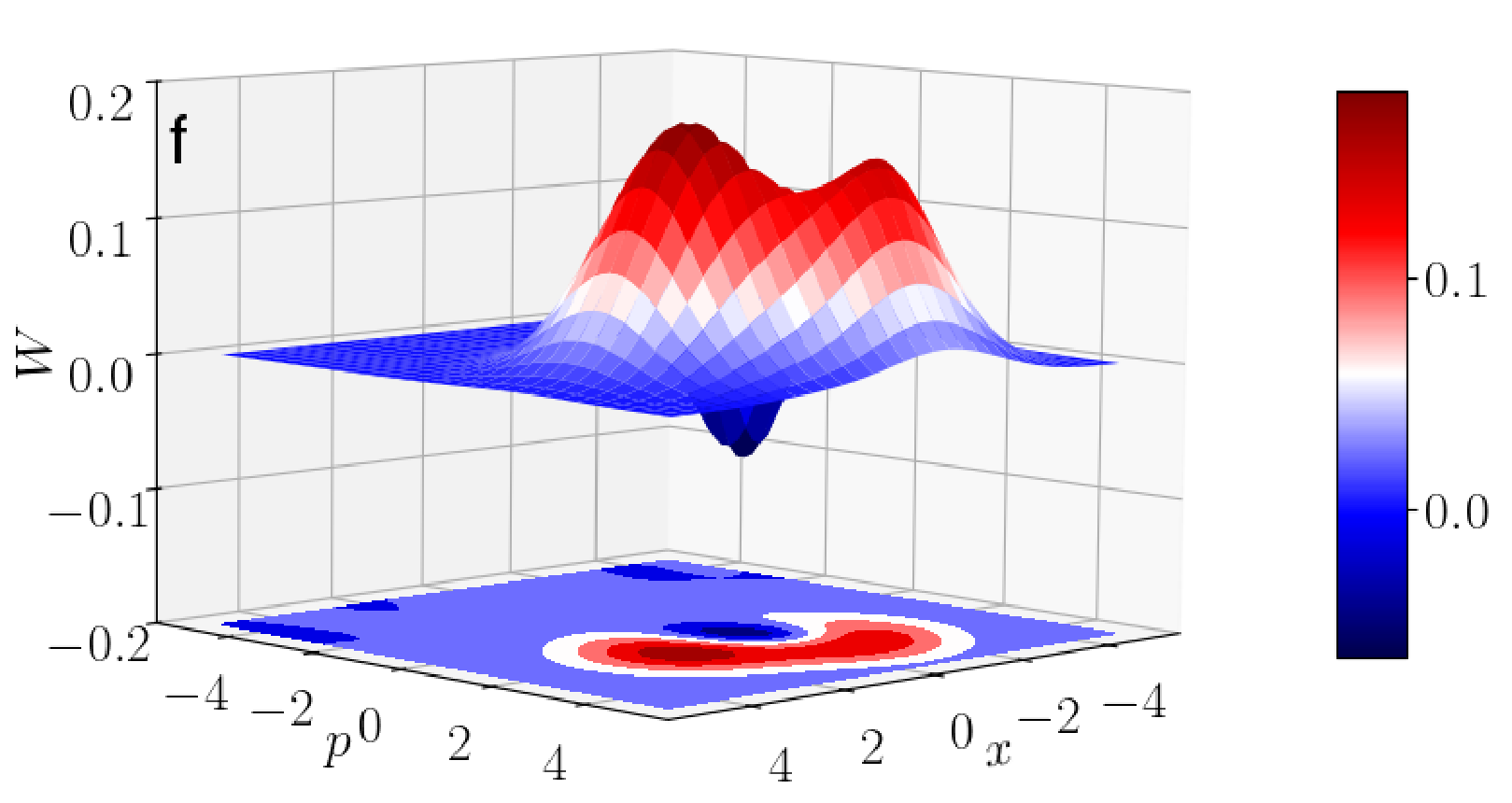}%
\caption{ The curvature–dependent Wigner functions versus $x$ and $p$ for $\alpha=1$, and $gt=3$,  at various curvature parameters: (a)~ $\lambda=0$, (b)~ $\lambda=0.1$, (c)~$\lambda=0.3$, (d)~ $\lambda=0.5$, (e)~$\lambda=0.7$, (f)~$\lambda=0.9$.}
\label{fig:5}
\end{figure*}
\subsection{  Curvature–dependent  Wigner function} \label{Wigner function}
The Wigner function provides a quasi-probability distribution in phase space derived from the quantum wave function. Although this function is real, it is not always positive across the entire phase space~\cite{weinbub2018recent}. The presence of negative regions in the Wigner function is a witness of the nonclassicality of a state, serving as a direct indicator of quantum interference and a measure of the state's quantumness~\cite{delfosse2015wigner,zurek2003decoherence}.
For a quantum state described by a density operator $\hat{\rho}$, the Wigner function in the phase space is given by
 $(\hbar=1)$:
\begin{eqnarray}\label{25}
W_{\psi}(x,p) &=&\frac{1}{2 \pi } \int^{\infty}_{-\infty} d\xi e^{-i p \xi} \langle x+\frac{\xi}{2}| \hat{\rho}| x-\frac{\xi}{2}| \rangle,
\end{eqnarray}
where $x$, and $p$ are position and momentum variables, respectively.  We start with the total system's density matrix $\hat{\rho}\equiv |\psi^{\lambda} (t) \rangle \langle \psi^{\lambda} (t)|$, trace over the atomic degrees of freedom to obtain the reduced field density matrix. For a pure state, the Wigner function~\eqref{25} simplifies to:
\begin{eqnarray}\label{26}
W_{\psi}(x,p) &=& \frac{1}{2 \pi } \int^{\infty}_{-\infty}d\xi e^{-i p \xi} \psi^{*}(x-\frac{\xi}{2}) \psi(x+\frac{\xi}{2}),
\end{eqnarray}
where $\psi^{\lambda}(x)$ is the wave function of our curvature–dependent atom–field system: $\langle x| \psi^{\lambda} (t) \rangle$,
and $|\psi^{\lambda} (t) \rangle $ is given by Eq.~\eqref{12}. By substituting $\psi^{\lambda} (x)$ in  Eq.~\eqref{26}, the curvature–dependent  Wigner function can be obtained as follows:
\begin{eqnarray}\label{27}
&&\hspace{-0.2cm} W^{\lambda}_{\psi}(x,p,t) = \frac{2 (x^{2}+p^{2})e^{(x^{2} + p^{2})}}{\pi} \sum^{\infty}_{n,m=0} (-1)^{n}2^{(m-n)/2} \sqrt{\frac{n!}{m!}} \nonumber  \\
&& \times  (x+ip)^{m-n} \Big(~|c^{\lambda}_{e,n}(t)|^{2}~ L^{m-n}_{n}
-  |c^{\lambda}_{g,n+1}(t)|^{2}~   L^{m-n}_{n+1}\Big),\,\,\,
\end{eqnarray}
where $L_{n}^{\alpha}(x)= \sum^{\infty}_{k=0} (-1)^{k} \binom{n+\alpha}{n-k} \frac{x^k}{k!}$ are generalized Laguerre polynomials~\cite{PhysRevA.54.4560}.

In Fig. \ref{fig:5},  we have plotted the curvature–dependent  Wigner function and it’s contour plot, versus $x$ and $p$ for different values of $\lambda$ with fixed parameters $gt = 3$ and $\alpha = 1$. In all cases, the presence of negative values clearly signifies the nonclassical nature of the field. In addition, increasing the curvature not only shifts the location of these negative regions but also influences their magnitudes.
Furthermore, the contour plots reveal asymmetrically declining peaks in phase space. In the center, distinct red and blue regions are apparent, each comprising multiple concentric circles that exhibit complete separation. Within their respective centers, the maximum positive and negative values of the Wigner function are observed. As $\lambda$ increases, the positive and negative peaks undergo a positional swap. Eventually, only one positive area and one negative area remain, which signifies a reduction in nonclassical statistical features as curvature increases.

\begin{figure*}[t]
\centering
{\includegraphics[width=0.7\columnwidth]{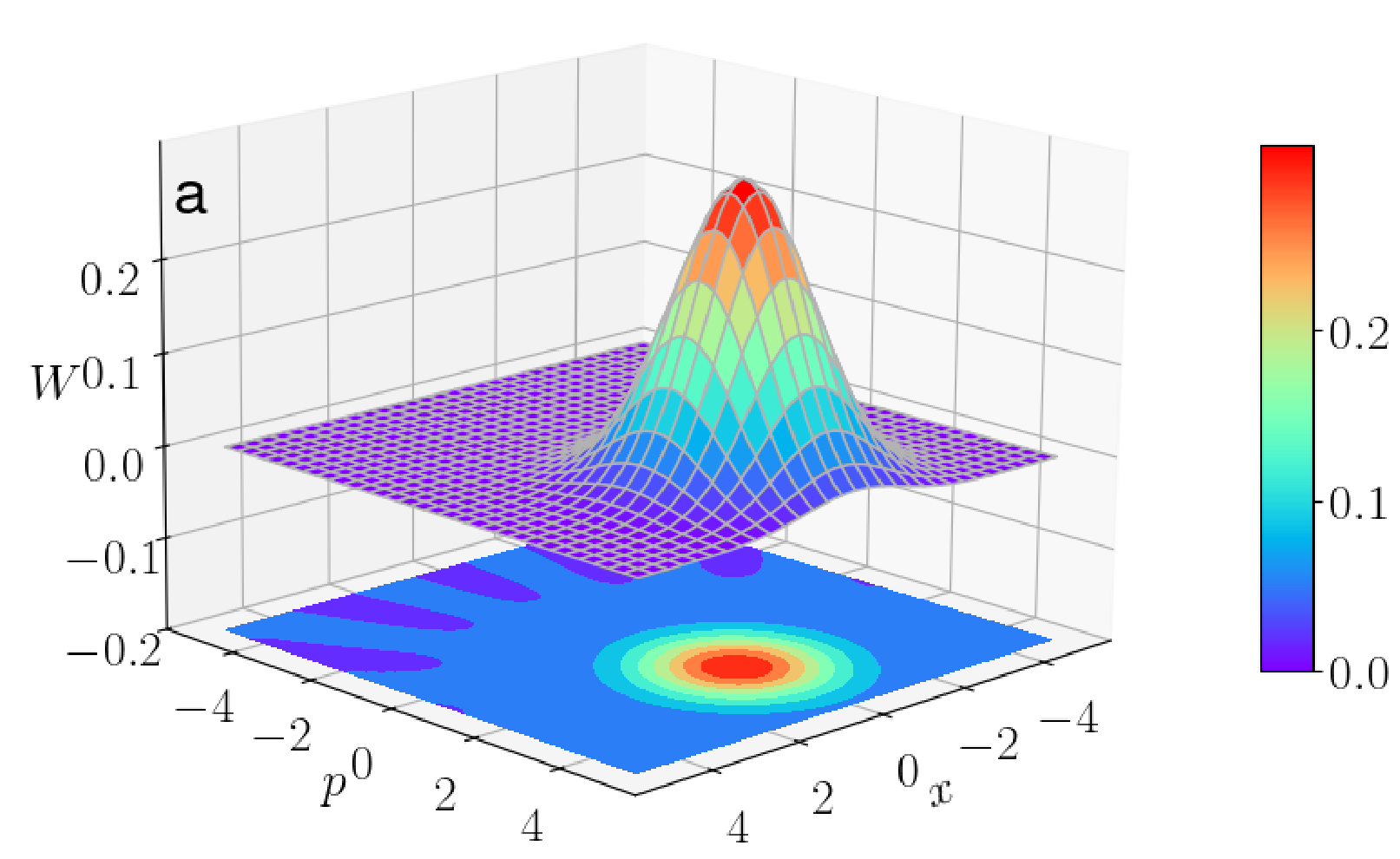}}%
{\includegraphics[width=0.7\columnwidth]{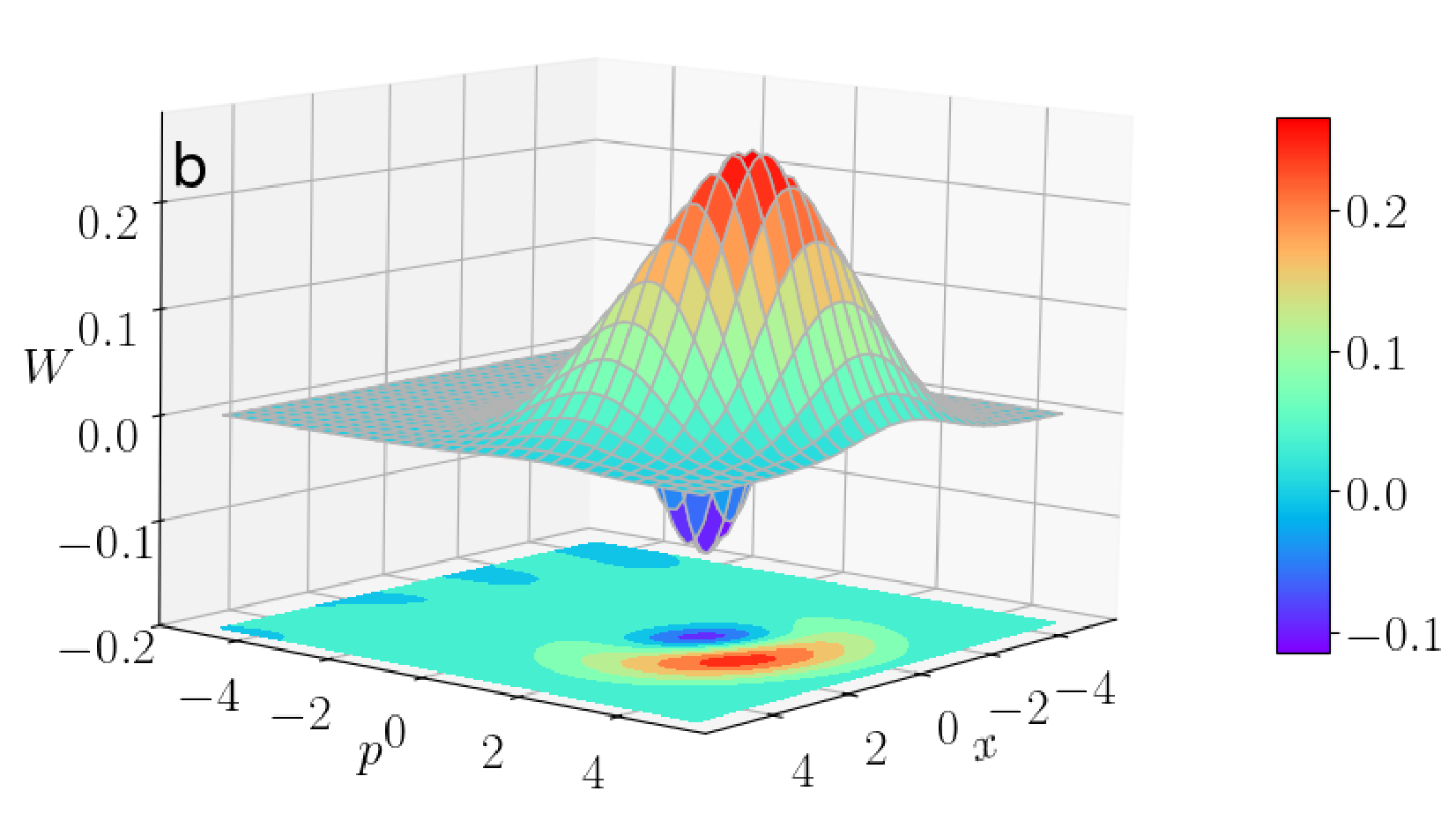}}%
{\includegraphics[width=0.7\columnwidth]{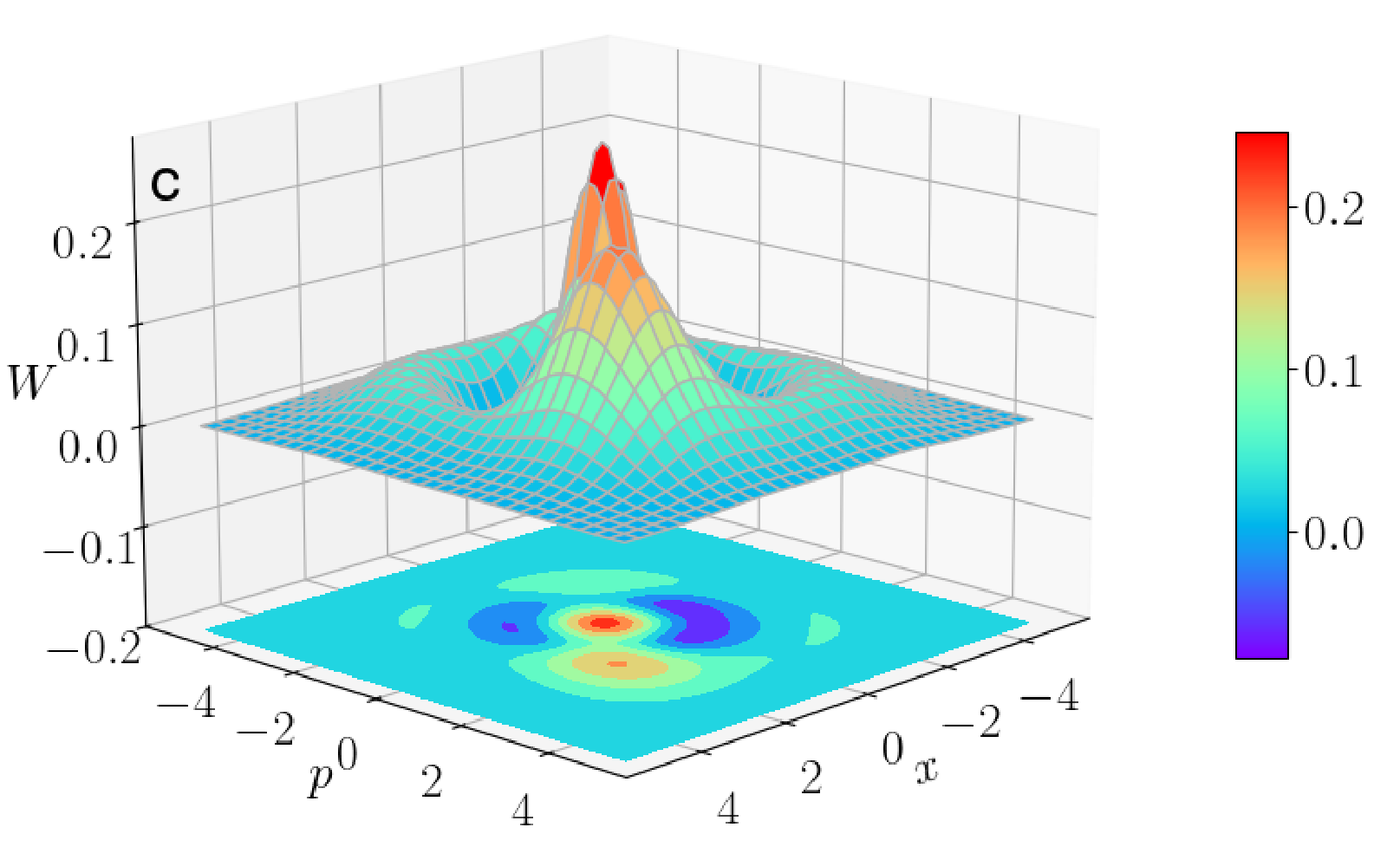}}%
\\
{\includegraphics[width=0.7\columnwidth]{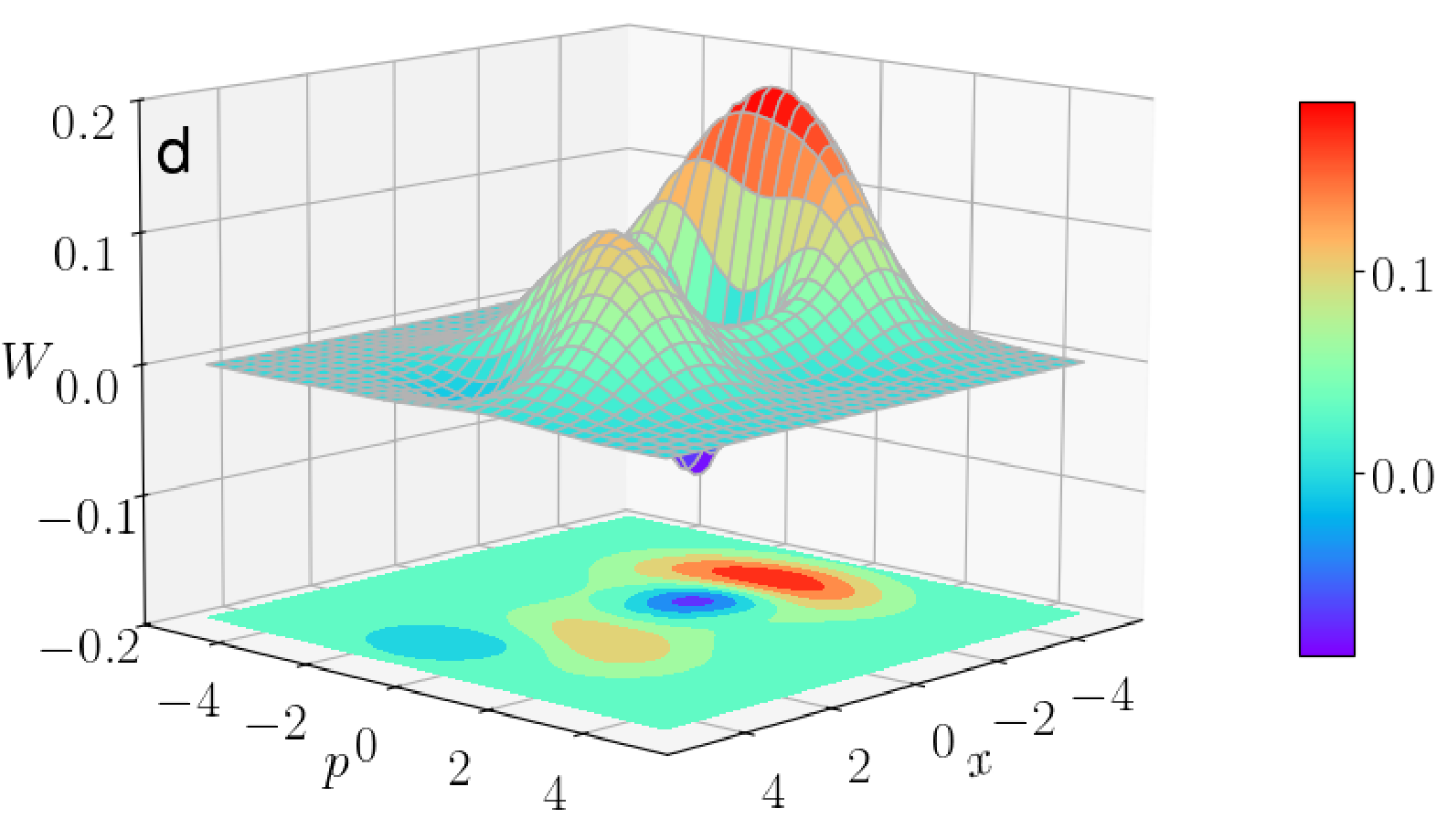}}%
{\includegraphics[width=0.7\columnwidth]{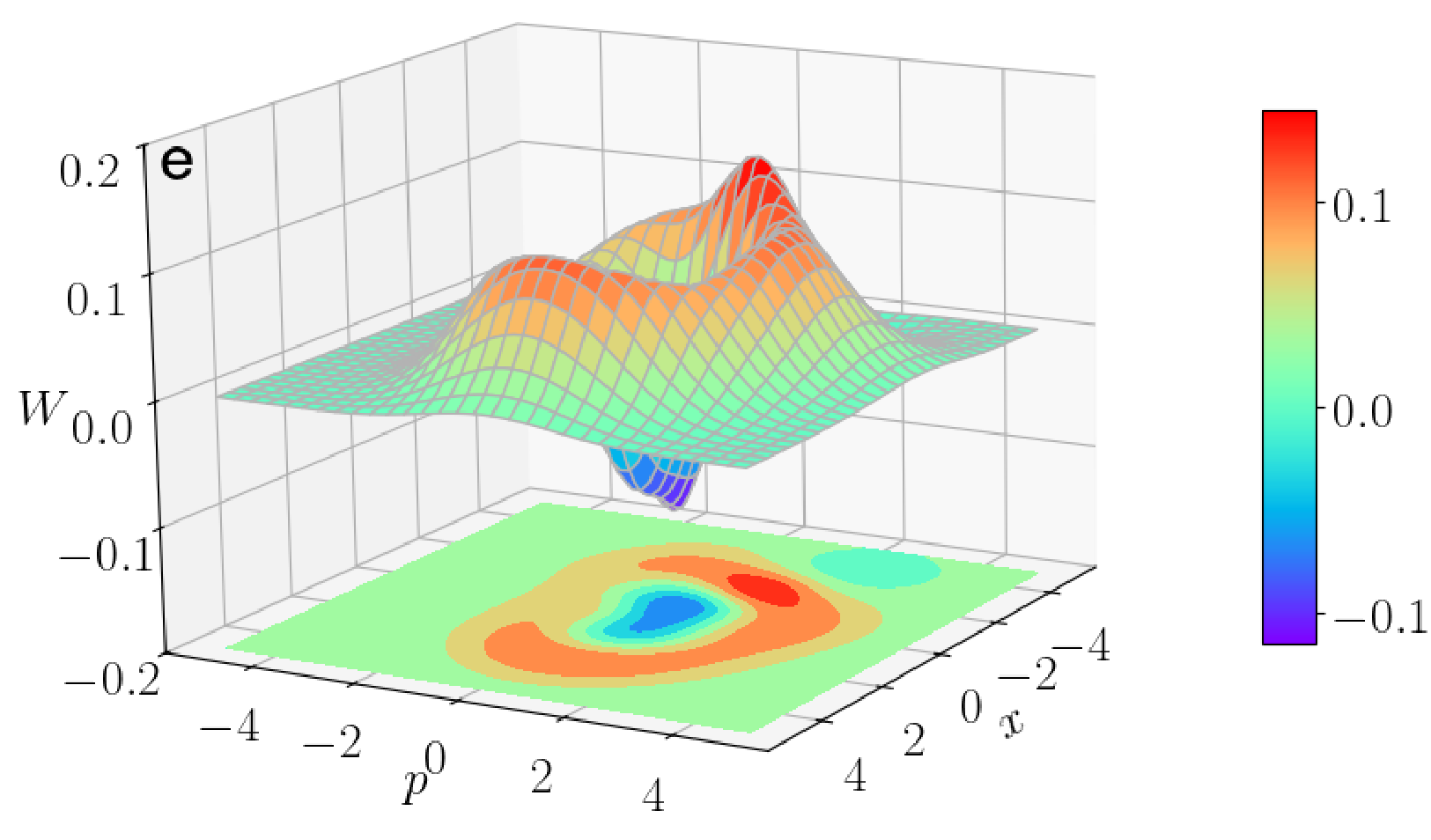}}%
{\includegraphics[width=0.7\columnwidth]{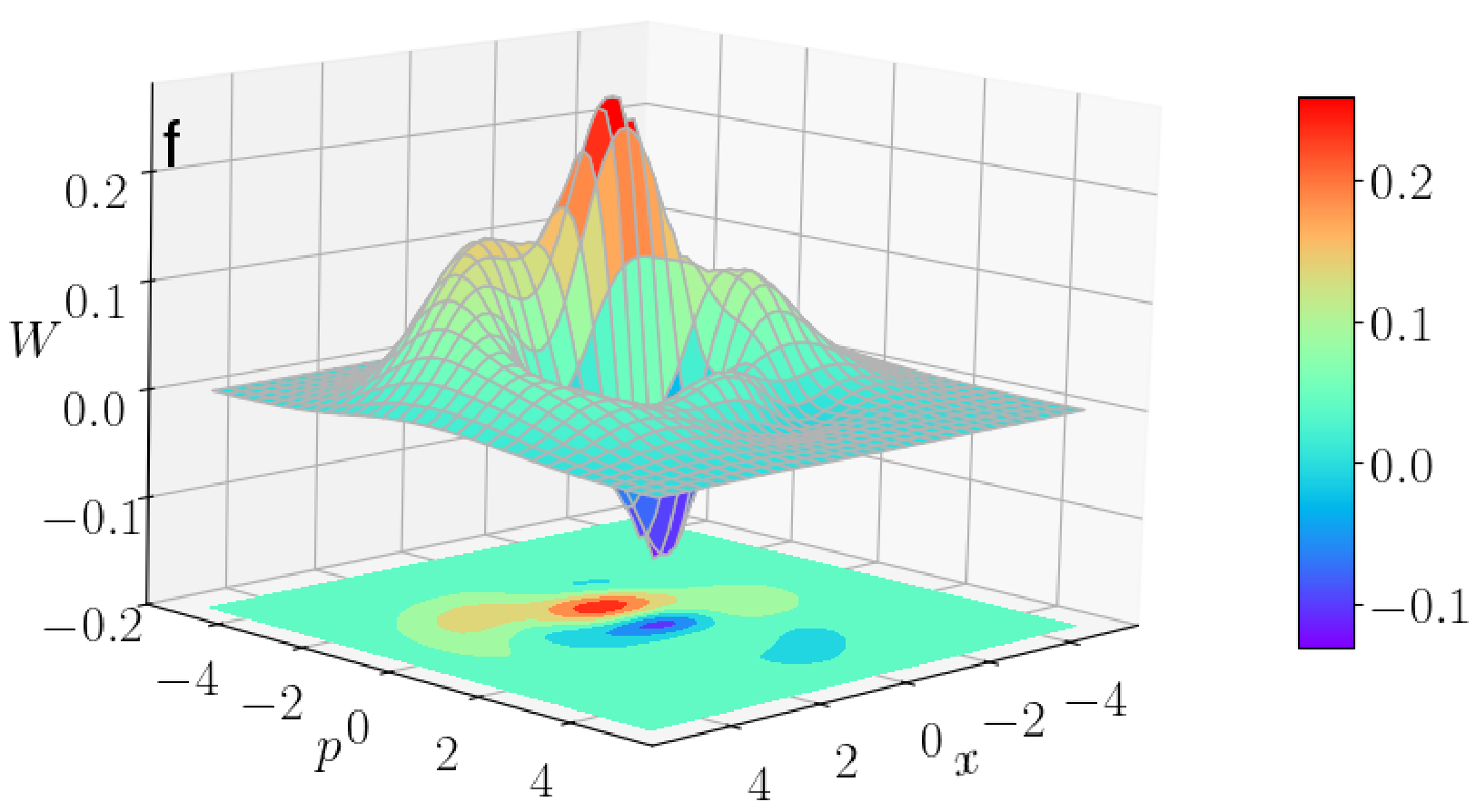}}%
\caption{The  curvature–dependent Wigner function versus $x$ and $p$  for $\lambda =0.3$, and $\alpha=1$, at different scaled times: (a)~$gt=0$, (b)~$gt=1$, (c)~$gt=3$, (d)~$gt=5$, (e)~$gt=7$, (f)~$gt=9$.}
\label{fig:6}
\end{figure*}

In Fig. \ref{fig:6}, we present the time evolution of the Wigner function for fixed curvature $\lambda =0.3$ with $\alpha=1$. At $gt = 0$, the field exhibits a Gaussian-like profile, closely resembling a coherent vacuum state, with a symmetric, positive peak centered near the origin. As time progresses, the Wigner function evolves from a unimodal shape at $ t=0$ to a more complex form. In some regions, it becomes negative and loses its Gaussian characteristics as a result of system interactions. The emergence of these negative regions illustrates the curvature-induced nonclassical effects within the system, manifested through a bifurcation pattern characterized by alternating positive and negative peaks.  These changes correspond to the dynamic probability redistribution between the excited state $|e,n\rangle$ and the ground state $|g,n+1\rangle$. At certain times, the field is more likely to be found in the ground state; while at others, the excited state becomes more probable.

\subsection{ Negativity of the Wigner Function}
It is well established that the negativity of the Wigner function is an indicator of non-classical behavior in quantum optics experiments. This negativity is employed to characterize states that cannot be described by classical physics, thereby revealing the quantum nature of a system ~\cite{chabaud2021witnessing,arkhipov2018negativity}.
The doubled volume of the integrated negative part of the Wigner function  of the state $|\psi^{\lambda}\rangle$ may be written as:
\begin{eqnarray}\label{34}
\delta^{\lambda} (\psi) &=& \int \int \Big[|W^{\lambda}_{\psi}(q,p,t)| - W^{\lambda}_{\psi}(q,p,t) \Big] dq dp \nonumber
\\\
&=& \int \int |W^{\lambda}_{\psi}(q,p,t)| dq dp -1.
\end{eqnarray}
By definition, the quantity $\delta^{\lambda} (\psi)$ vanishes for coherent and squeezed vacuum states, whose Wigner functions remain strictly non-negative~\cite{kenfack2004negativity,taghiabadi2016revealing,schleich2015quantum}.
\begin{figure}
\centering
\includegraphics[width=3 in]{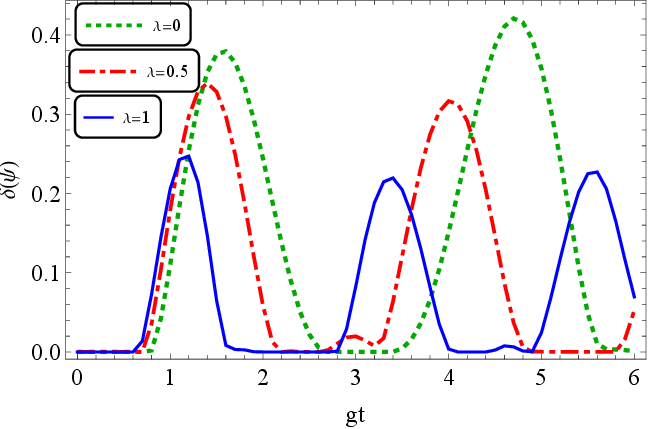}
\caption{Negativity of the Wigner function versus $gt$  for $\alpha=0.5$. The dotted green curve to $\lambda=0$, the dotted-dashed red curve to  $\lambda=0.5$  and the solid blue curve corresponds $\lambda=1$. }
 \label{fig:7}
\end{figure}

In Fig.~\ref{fig:7}, we have  plotted the negativity of the Wigner function as a function of $gt$.  It is seen that $\delta^{\lambda} (\psi)$ is initially zero, but increases over time until it reaches a maximum and subsequently decreases. Additionally, by increasing spatial curvature, the onset of Wigner function negativity occurs in shorter time intervals. Moreover, the maximum value of $\delta^{\lambda} (\psi)$ decreases with increasing $\lambda$, indicating a curvature-dependent suppression of nonclassicality.
%
\subsection{Quantum Entanglement and Curvature–dependent Entropy}
The von Neumann entropy plays a fundamental role in quantifying quantum entanglement and nonclassicality~\cite{ohya2004quantum}.
Entanglement is widely recognized as one of the most non-classical characteristic of a quantum state ~\cite{bianchini2014entanglement}. The von Neumann entropy is defined as:
\begin{equation}\label{288}
S_{N}= - k_{B} Tr{\hat{\rho}_{a,f} \ln \hat{\rho}_{a,f}}.
\end{equation}
where $k_{B}$ is Boltzmann's constant, and $\hat{\rho}_{a,f}$ is the joint density operator of the atom-field system.
The reduced atomic density operator $\hat{\rho}_{a}$ is obtained by tracing out the field degrees of freedom:
\begin{eqnarray}\label{27}
\hat{\rho}_{a}(t)&=& \Sigma^{\infty} _{n=0} \Big[|c^{\lambda}_{e,n}|^{2} |e\rangle \langle e| +
 c^{\lambda}_{e,n}c^{*\lambda}_{g,n+1} |e\rangle \langle g| \nonumber\\\ &+& c^{\lambda}_{g,n+1}c^{*\lambda}_{e,n} |g\rangle \langle e| + |c^{\lambda}_{g,n+1}|^{2} |g\rangle \langle g| \Big]
\end{eqnarray}
We can express the atomic entropy in terms of the eigenvalues of the reduced atomic density operator as follows~\cite{buvzek1995quantum}:
\begin{equation}\label{32}
S_{a}=- \sum_{m=1,2} \rho_{m}(t) \ln \rho_{m}(t),
\end{equation}
where $\rho_{m}(t)$ are the eigenvalues of the reduced atomic density  matrix $\rho_{a}$.
\begin{figure}
\centering
\includegraphics[width=2.8 in]{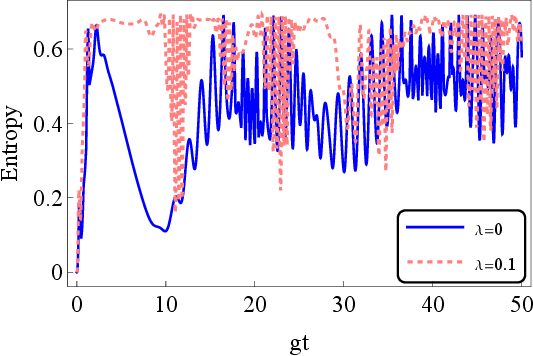}
\caption{ The time evolution of the curvature–dependent entropy versus scaled time, $gt$, assuming a coherent state as the initial field with $\alpha = 3$. The solid blue curve corresponds $\lambda= 0$, while the dashed red curve represents $\lambda=0.1$. }
 \label{fig:11}
\end{figure}
\begin{figure}
\centering
\includegraphics[width=2.8 in]{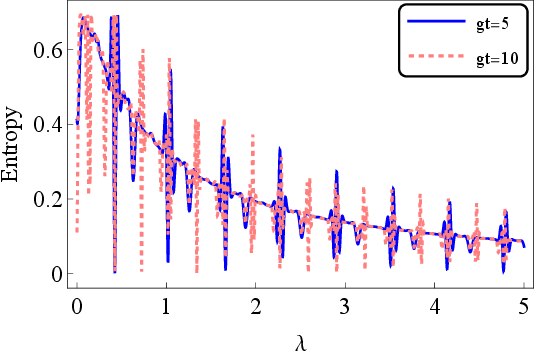}
\caption{ The time evolution of the curvature-dependent entropy as a function of $\lambda$ is shown for a specific time with $\alpha = 3$. The solid blue curve corresponds to $gt = 5$, while the dashed red curve represents $gt = 10$.}
 \label{fig:12}
\end{figure}
In Fig. \ref{fig:11}, the time evolution of the  entropy is plotted. We observe that the curvature-dependent von Neumann entropy initially increases and then decreases as $\lambda$ grows.
In addition, rapid oscillations are observed for all values of $\lambda$. Moreover, increasing the curvature reduces the time interval between collapse and revival phenomena.

Fig. \ref{fig:12} shows the dynamical behavior of the entropy versus $\lambda$.
It reveals that after an initial rise, the entropy decreases with increasing curvature parameter $\lambda$, highlighting a curvature-induced modulation of entanglement dynamics.
 %
 \section{SUMMARY AND CONCLUDING REMARKS}\label{SUMMARY}
In summary, by solving the Schrodinger equation for the $\lambda$–dependent JCM under the rotating wave approximation in the interaction picture, based on the quantum oscillators on a circle, the time evolution of the atom-field system is investigated. This allowed us to investigate the effects of spatial curvature, characterized by the parameter $\lambda$, on various dynamical properties of both the atomic system and the field.
The spatial curvature plays an important role in determining the nonclassical features of the  $\lambda$-dependent JCM. Physically, the parameter $\lambda$ can be interpreted as a measure of geometric-induced nonlinearity. As $\lambda$ varies, the nonlinearity of the system changes correspondingly, leading to an enhancement or suppression of its nonclassical properties.

The oscillators on a circle and their associated curvature-dependent states, exhibit various nonclassical properties, such as sub-Poissonian statistics and negativity in the Wigner function.
We found that spatial curvature serves as a controlling factor for these nonclassical properties: specifically, increasing the curvature tends to reduce the degree of nonclassicality exhibited by the system.

\begin{acknowledgements}
The author would like to express his sincere thanks to  support for this work by the Office of Graduate Studies and Research of Isfahan for their support. We would also like to extend our thanks to Shahrekord University for  their assistance.
\end{acknowledgements}

\bibliographystyle{apsrev4-2}
\bibliography{WingerFbib}

\end{document}